\documentclass[a4paper]{aa}
\usepackage{graphicx,float}

\def \src {X\thinspace1254$-$690}
\def \degmark{^\circ}

\def \ergsec{\hbox{erg s$^{-1}$}}
\def \hcm {\hbox {\ifmmode $ atom cm$^{-2}\else atom cm$^{-2}$\fi}}
\def \arcmin {\hbox{$^\prime$}}
\def \arcsec {\hbox{$^{\prime\prime}$}}

\def\approxgt{\mathrel{\hbox{\rlap{\lower.55ex \hbox {$\sim$}}
        \kern-.3em \raise.4ex \hbox{$>$}}}}
\def\approxlt{\mathrel{\hbox{\rlap{\lower.55ex \hbox {$\sim$}}
        \kern-.3em \raise.4ex \hbox{$<$}}}}

\begin{document}

\title{Discovery of narrow X-ray absorption features 
from the low-mass X-ray binary \src\ with XMM-Newton}

\author{L. Boirin \and A. N. Parmar
}
\offprints{L. Boirin, \email{lboirin@rssd.esa.int}}

\institute{
       Astrophysics Division, Research and Scientific Support 
       Department of ESA, ESTEC,
       Postbus 299, NL-2200 AG Noordwijk, The Netherlands
}
\date{Received 22 April 2003 / Accepted 19 June 2003}

\authorrunning{L. Boirin \& A. N. Parmar}

\titlerunning{XMM-Newton observations of \src}

\abstract{We report on two XMM-Newton observations of the low-mass
X-ray binary \src. During an XMM-Newton observation of the low-mass
X-ray binary in 2001 January a deep X-ray dip was seen while in a
second observation one year later no dips were evident.  The
0.5--10~keV EPIC spectra from both non-dipping intervals are very
similar being modeled by a disk-blackbody and a power-law continuum
with additional structure around 1~keV and narrow absorption features
at 7.0~keV and 8.2~keV which are identified with the K$\alpha$ and
K$\beta$ absorption lines of Fe\,{\sc xxvi}.  The low-energy structure
may be modeled as a 175~eV ($\sigma$) wide emission line at
$\sim$0.95~keV. This feature is probably the same structure that was
modeled as an absorption edge in an earlier BeppoSAX observation.  The
absorption line properties show no obvious dependence on orbital phase
and are similar in both observations suggesting that the occurrence of
such features is not directly related to the presence of dipping
activity.  Narrow Fe absorption features have been observed from the
two superluminal jet sources GRO\,J1655$-$40 and
GRS\thinspace1915$+$105, and the four low-mass X-ray binaries
GX\thinspace13+1, MXB\thinspace1658$-$298, X\thinspace1624$-$490 and
\src. Since the latter 3 sources are dipping sources, which are
systems viewed close to the accretion disk plane, and the two
microquasars are thought to be viewed at an inclination of
$\sim$70$\degmark$, this suggests that these features are more
prominent when viewed at high-inclination angles.  This, together with
the lack of any orbital dependence, implies a cylindrical geometry for
the absorbing material.

  } \maketitle

\section{Introduction}
\label{sect:intro}

During EXOSAT observations in 1984 and 1985 the low-mass X-ray binary
(LMXRB) source \src\ exhibited irregular dips in X-ray intensity that
repeated every $3.88 \pm 0.15$~hr (Courvoisier et al.~\cite{c:86}).
The duration of each dip was $\sim$0.8~hr with a mean reduction in
1--10~keV intensity of $\sim$95\%.  From optical V-band observations
of the 19th magnitude companion Motch et al.~(\cite{m:87}) determined
an optical period of $3.9334 \pm 0.0002$~hr, consistent with the mean
X-ray dip recurrence interval.  The dips are due to obscuration in the
thickened outer regions of an azimuthally structured accretion disk
(White \& Swank~\cite{ws:82}).  The dips were still present during
{\it Ginga} observations in 1990 (Uno et al.~\cite{u:97}), but were
not detected by RXTE in 1997 (Smale \&
Wachter~\cite{sw:99}). Simultaneous optical observations revealed that
the mean optical magnitude was unchanged, but that the amplitude of
the optical variability had declined (Smale \& Wachter~\cite{sw:99}).
The dips were also absent during a BeppoSAX observation in 1998 (Iaria
et al.~\cite{i:01}).  A likely explanation for the absence of dips is
that the vertical structure in the outer region of the accretion disk
had decreased in size so that the central X-ray source could be viewed
directly (Smale \& Wachter~\cite{sw:99}).  The dips had re-appeared by
the time of a RXTE observation in 2001 May, but were again not present
in 2001 December (Smale et al.~\cite{s:02}).  The 0.1--100~keV
BeppoSAX spectrum of \src\ may be modeled by the combination of a
multicolor disk-blackbody with an inner temperature of $\sim$0.85~keV
and a Comptonized component with an electron temperature of
$\sim$2~keV and an optical depth of $\sim$19.  A hard excess is
visible around 20~keV which may be accounted for using a
bremsstrahlung model with a temperature of $\sim$20~keV (Iaria et
al.~\cite{i:01}).  There was also evidence in the BeppoSAX observation
for an absorption edge at $\sim$1.27~keV with an optical depth of
$\sim$0.15.  Iaria et al.~(\cite{i:01}) propose that the Comptonized
component could originate in a spherical cloud, or boundary layer,
surrounding the neutron star while the bremsstrahlung component
probably originates in an extended accretion disk corona with a radius
of $10^{10}$~cm and a mean electron density of $\sim$$1.7 \,
10^{14}$~cm$^{-3}$.

\begin{figure*}[t!]
 \hbox{\hspace{-0.7cm}
 \includegraphics[width=7.35cm,angle=90]{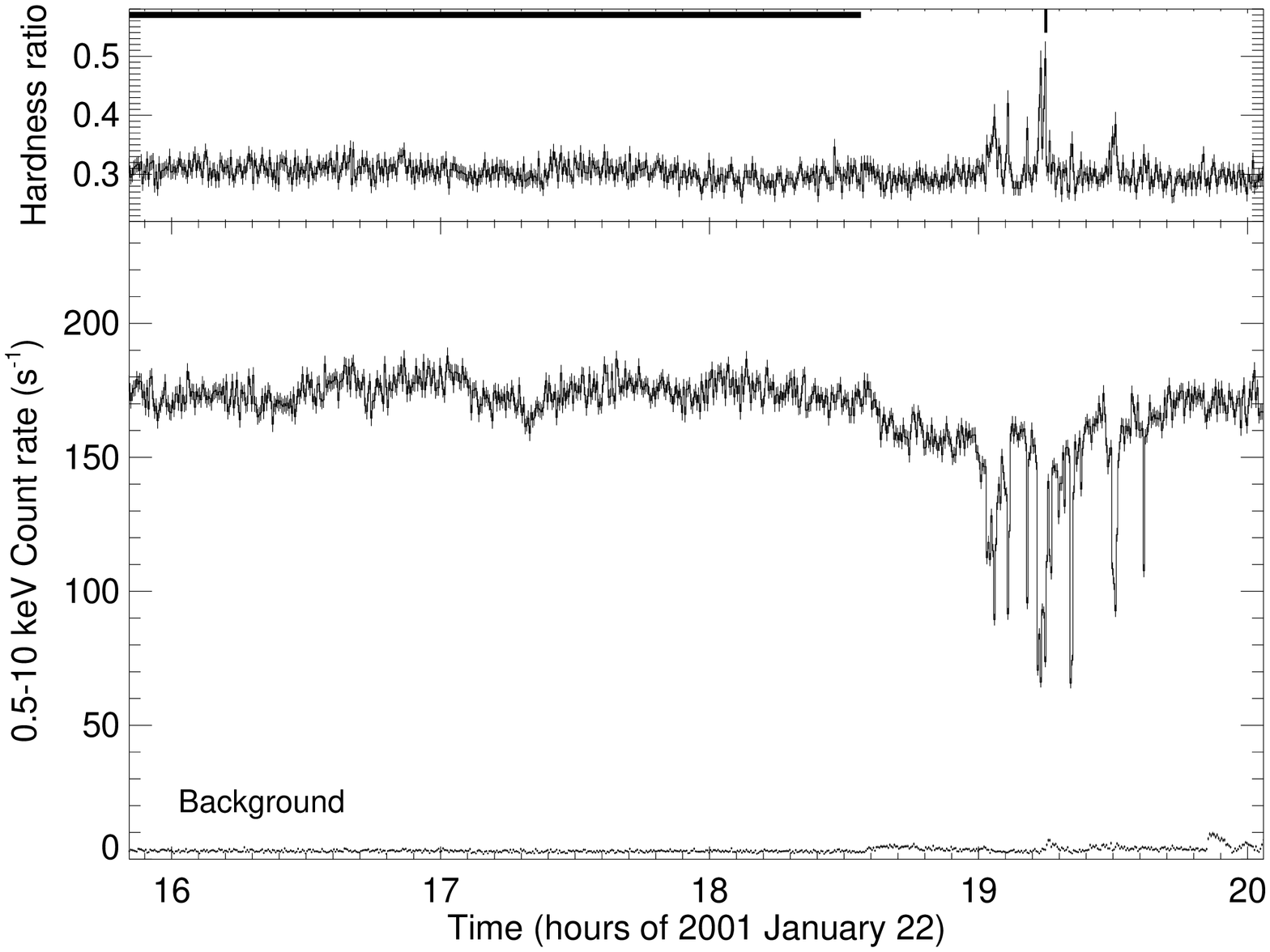}
 \hspace{-0.7cm}
 \includegraphics[width=7.35cm,angle=90]{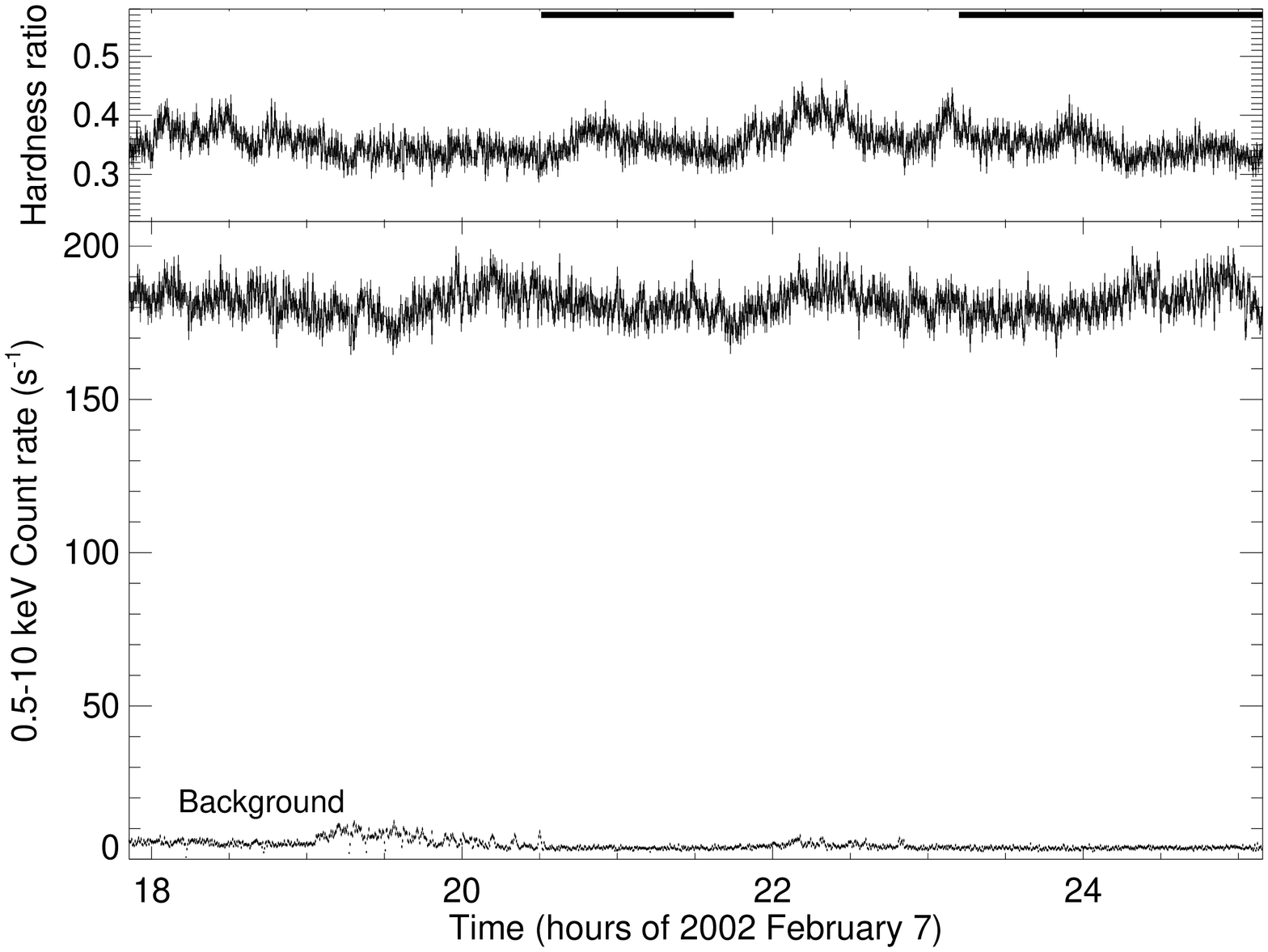}}
 \caption[]{The EPIC PN 0.5--10 keV background subtracted lightcurves
 of \src\ with a binning of 20~s (lower panels) showing a deep dip
 during the 2001 observation and the lack of dipping activity during
 the 2002 observation. The upper panels show the hardness ratios
 (counts between 3--10~keV divided by those between 0.5--3~keV) also
 with a binning of 20~s. The 0.5--10~keV background counts taken from
 regions adjacent to the source are also shown. The thick horizontal
 lines indicate the intervals included in the spectral analysis. The
  tick mark flags the estimated dip center time chosen as phase 0.}
 \label{fig:lightcurve}
\end{figure*}

\begin{figure}
\hbox{\hspace{-0.5cm}
\includegraphics[width=7.0cm,angle=90]{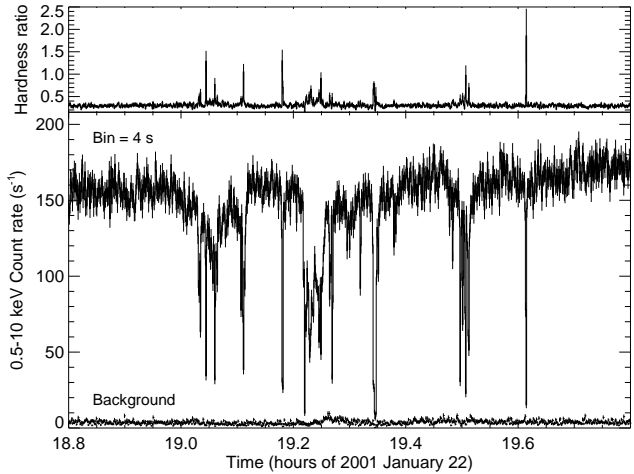}}
\caption[]{A 0.5--10~keV PN background subtracted lightcurve of the
dip seen in the 2001 January observation plotted with a time
resolution of 4~s showing the rapid intensity variability during the
dip.  The upper panel shows the hardness ratio (counts between
3--10~keV divided by those between 0.5--3~keV).}
\label{fig:zoom_dip}
\end{figure}

The PN camera (Str\"uder et al. \cite{st:01}) of the European Photon
Imaging Camera (EPIC) on-board XMM-Newton (Jansen et al. \cite{j:01}),
with its good high-energy sensitivity and spectral resolution, is
proving to be very successful in discovering narrow absorption
features from highly ionized Fe and other metals in the spectra of
LBXRBs. 

 Narrow X-ray absorption lines were first detected from the
superluminal jet sources GRO\,J1655$-$40 (Ueda et al.~\cite{u:98}) and
GRS\,1915+105 (Kotani et al.~\cite{k:00}), and it has been suggested
that these features are related to the jet formation mechanism. With
the detection of X-ray absorption lines in GX\thinspace13+1,
MXB\thinspace1658$-$298, X\thinspace1624$-$490 and now \src, this now
appears unlikely. X-ray absorption lines were seen from
GX\thinspace13+1 using ASCA by Ueda et al.~(\cite{u:01}) who detected
a narrow absorption feature at 7.01~keV which they interpreted as
resonant scattering of the K$\alpha$ line from H-like Fe.  XMM-Newton
observations revealed an even more complex picture for GX\,13+1 with
narrow absorption features identified with the K$\alpha$ and K$\beta$
transitions of He- and H-like iron (Fe\,{\sc xxv} and Fe\,{\sc xxvi})
and H-like calcium (Ca\,{\sc xx}) K$\alpha$ detected (Sidoli et
al.~\cite{si:02}).  There is also evidence for the presence of a deep
Fe~{\sc xxv} absorption edge at 8.83~keV and a broad emission feature
at around 6.4~keV.  XMM-Newton observations of the eclipsing and
dipping LMXRB MXB\thinspace1658$-$298 revealed narrow absorption
features identified with O\,{\sc viii}~K$\alpha$, K$\beta$, and
K$\gamma$, Ne\,{\sc x} K$\alpha$, Fe\,{\sc xxv}~K$\alpha$, and
Fe\,{\sc xxvi}~K$\alpha$ together with a broad Fe emission feature at
6.47~keV (Sidoli et al.~\cite{si:01}).  Another LMXRB dip source,
X\thinspace1624$-$490, also displays K$\alpha$ absorption lines
identified with Fe\,{\sc xxv} and Fe\,{\sc xxvi} as well as fainter
absorption features tentatively identified with Ni\,{\sc xxvii}
K$\alpha$ and Fe\,{\sc xxvi} K$\beta$.  A broad emission feature at
6.58~keV is also evident (Parmar et al.~\cite{p:02}).  The properties
of the absorption features in the 3 LBXRBs show no obvious dependence
on orbital phase, except during a dip from X\thinspace1624$-$490 where
there is evidence for the presence of additional cooler material.
Here we report the discovery of narrow X-ray absorption features from
highly ionized Fe in the XMM-Newton spectrum of \src.  As we discuss
below, the presence of X-ray absorption features is probably related
to the viewing angle of the system.

\begin{figure*}
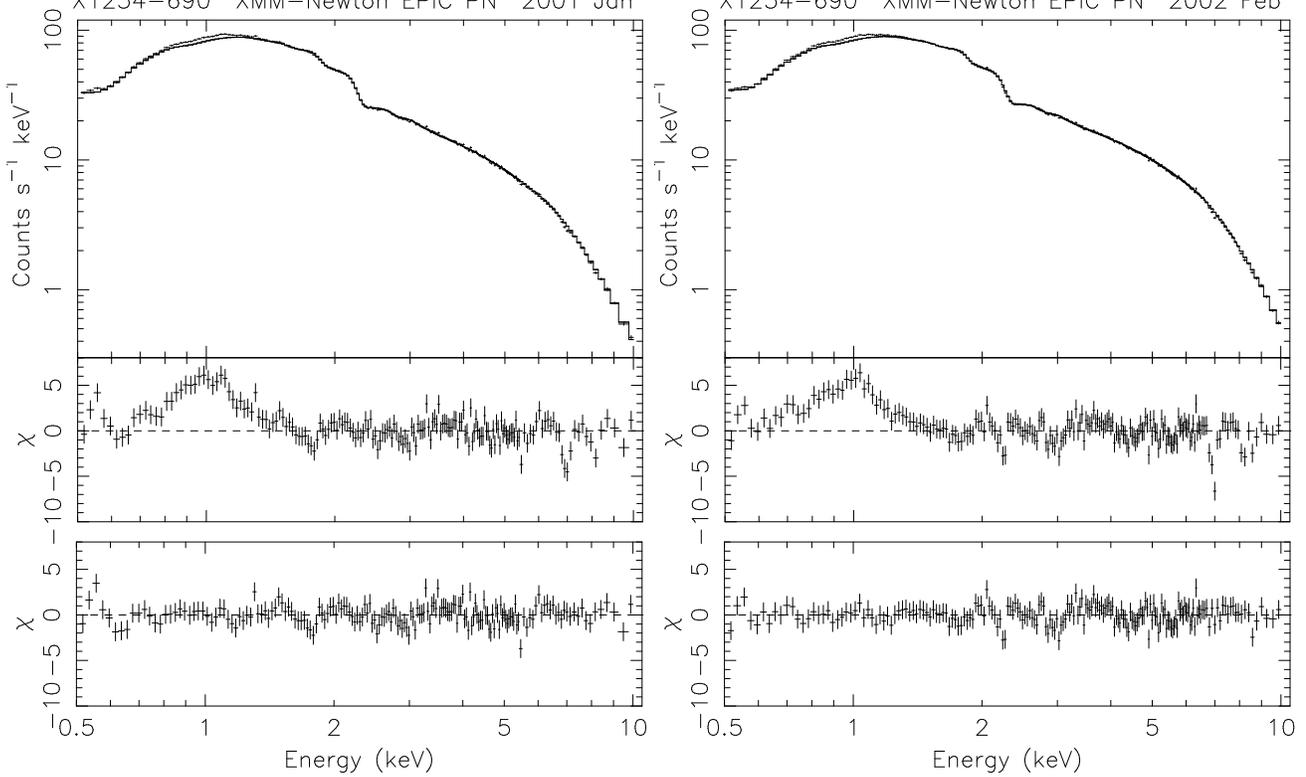

  \hbox{\hspace{0.0cm}
   \includegraphics[height=8.5cm,angle=-90]{H4477F3a.ps}
   \includegraphics[height=8.5cm,angle=-90]{H4477F3b.ps}}
  \vspace{-0.1cm}
  \hbox{\hspace{0.15cm}
   \includegraphics[height=8.4cm,angle=-90]{H4477F3c.ps}
   \hspace{0.0cm}
   \includegraphics[height=8.4cm,angle=-90]{H4477F3d.ps}}
  \caption[]{The upper panels shows the PN spectra of the \src\ 
             persistent emission
             and the best-fit disk-blackbody and power-law 
             continuum models. The residuals (middle panels) reveal the
             presence of a broad emission feature centered at 
             $\sim$0.93~keV, together with narrow absorption 
             features at 6.96~keV and 8.16~keV.
             The lower panels show the residuals when these features
             are included in the spectral model.}
  \label{fig:spectrum}
\end{figure*}

\section{Observations}
\label{sect:obs}

%Due to the high count rate, only Timing mode provides useful spectral
%information in the MOS, we concentrate on the analysis of PN data, and
%use the lower count rate MOS data to check for consistency.

The XMM-Newton Observatory includes three 1500~cm$^2$ X-ray telescopes
each with an EPIC at the focus.  Two of the EPIC imaging spectrometers
use MOS CCDs (Turner et al.  \cite{t:01}) and one uses a PN CCD.  The
region of sky containing \src\ was observed by XMM-Newton between 2001
January 22 15:49 UT to 20:03~UT and again between 2002 February 07
17:32 UT to February 08 01:08~UT.  In order to minimize the effects of
pile-up the PN was operated in Timing mode during both observations.
In this mode only one central CCD is read out with a time resolution
of 0.03~ms.  This provides a one dimensional (4.4\arcmin\ wide) image
of the source with the second dimension being replaced by timing
information.  The faster CCD readout results in a much higher count
rate capability of 1500~s$^{-1}$ before charge pile-up become a
serious problem.  One MOS camera was operated in Small Window mode and
the other was operated in Timing mode. As the PN effective area is a
factor 5 higher than the MOS one at around 7~keV where the absorption
features are discovered, we concentrate here on the analysis of PN
data. The PN camera was operated with thin filters during both
observations.

Raw data products were extracted from the public XMM-Newton archive
and processed using version 5.4.1 of the Science Analysis Software
(SAS). Only single and double X-ray events corresponding to patterns
0--4 were selected.  Known hot, or flickering, pixels and electronic
noise were rejected using the SAS.  Both observations contain high
background intervals due to solar activity. These intervals were
excluded for the spectral analysis.

Source events were extracted from a column 70\arcsec\ wide centered on
\src.  Background events were obtained from a column of the same
width, but centered 100\arcsec\ from \src. The background events will
be contaminated by a small number of source events due to the extent
of the point spread function.  However, background subtraction is not
critical for such a bright source.  The mean count rates during both
observations of 175 and 185~s$^{-1}$ are well below the value
(1500~s$^{-1}$) where source pile-up becomes a problem in PN Timing
mode.

\section{Results}
\subsection{X-ray lightcurves}

The 0.5--10~keV \src\ background subtracted lightcurves obtained from
the PN during the two observations are shown in
Fig.~\ref{fig:lightcurve} with a binning of 20~s. The upper panels
show the PN hardness ratios (counts in the energy range 0.5--10~keV
divided by those between 0.5--3~keV) also with a binning of 20~s. The
source intensity was less during the 2001 observation with a mean
count rate of 185~s$^{-1}$ compared to 175~s$^{-1}$ during the 2002
February observation.  During the 2001 January observation a deep dip
occurs between 19.0 hrs and 19.6~hrs on 22 January 2001 with irregular
intensity variability and an associated increases in hardness ratio.
During the dip the count rate drops as low as 70~s$^{-1}$ when plotted
at 20~s time resolution.  Fig.~\ref{fig:zoom_dip} shows an expanded
view of the dip with a time resolution of 4~s.  At times, the dipping
in almost total in the 0.5--10~keV energy range when plotted with this
time resolution.  In the 2002 observation, the 0.5--10~keV source
count rate varies between $\sim$170-200~s$^{-1}$ with flare-like
variability, but there is no evidence for any dipping activity.

\begin{table*}
\begin{center}
\caption[]{Best-fits to the 0.5--10~keV XMM-Newton PN persistent
emission spectra of \src\ using a disk-blackbody model with an inner
temperature $KT_{\rm in}$ and normalization $k$ given by $(R_{\rm
in}/d)^2$ where $R_{\rm in}$ in the inner radius in km and $d$ the
distance in units of 10~kpc and a power-law with a photon index,
$\alpha$ together with an emission and two absorption lines.  }
\begin{tabular}{llcc}
\hline
\noalign {\smallskip}
Component & Parameter&2001 January              &  2002 February \\
\noalign {\smallskip}
\hline
\noalign {\smallskip}
& $N{\rm _H}$  ($10^{21}$~atom cm$^{-2}$) & $3.2 \pm 0.1$ &  $3.1 \pm 0.1$ \\

&$L$ (\ergsec, at 10 kpc)    & $1.1 \, 10^{37}$  & $1.0 \, 10^{37}$  \\

&$\chi ^2$/dof        & 264.6/225   & 244.6/225 \\

Disk-blackbody &$kT_{\rm in}$ (keV)   &$1.97 \, ^{+0.04} _{-0.07}$  
& $2.31 \pm 0.03 $ \\
          &$k$  & $1.41 \pm 0.10$  & $1.06 \pm 0.04$ \\

Power-law&$\alpha$    & $2.20 \pm 0.09$ & $2.34  \pm 0.09$ \\
          &1 keV normalization& $0.106 \pm 0.006$  & $0.099 \pm 0.005$ \\

Emission line 
& $E_{{\rm line}}$ (keV)     & $0.96\, ^{+0.04} _{-0.06}$ 
& $0.93 \, ^{+0.05} _{-0.09}$ \\

& $\sigma$ (eV)              & $175 \, ^{+75} _{-50}$   & $175 \pm 65$    \\
&EW              (eV)        & $32 \pm 15$   & $29 \pm 13$ \\

Fe\,{\sc xxvi} K$\alpha$ abs
& $E_{{\rm line}}$ (keV)     &$6.95 \pm 0.03$    & $6.96 \pm 0.04$ \\
feature&$\sigma$ (eV)                  &$<$120  &  $<$95    \\
&EW              (eV)         &$-27 \, ^{+11} _{-8}$  
& $-21 \, ^{+8} _{-5}$ \\

Fe\,{\sc xxvi} K$\beta$ abs
& $E_{{\rm line}}$ (keV)      &$8.20 \, ^{+0.05} _{-0.10}$  
& $8.16 \pm 0.06$ \\
feature&$\sigma$ (eV)          & $<$170     &   $<$80    \\
&EW              (eV)         & $-17 \pm 9$ & $-16 \pm 9$ \\

\noalign {\smallskip}                       
\hline
\label{tab:spectrum}
\end{tabular}
\end{center}
\end{table*}

\begin{table*}
\begin{center}
\caption[]{Properties of the Fe\,{\sc xxvi} K$\alpha$ absorption
feature in three segments of the persistent emission during the 2001
observation. The time reference used for phase 0 is 19.25~h of 2001 January 22 and the period is 3.88~h.}
\begin{tabular}{llccc}
\hline
\noalign {\smallskip}
 & & Segment 1 & Segment 2 & Segment 3 \\
\noalign {\smallskip}
\hline
\noalign {\smallskip}

\multicolumn{2}{l}{Phase range} & 0.12-0.35 & 0.35-0.59 & 0.59-0.82 \\

\multicolumn{2}{l}{Fe\,{\sc xxvi} K$\alpha$ abs feature}\\
&$E_{{\rm line}}$ (keV)     &$6.93 \pm 0.09$    & $6.95 \, ^{+0.03} _{-0.06}$ & $6.97 \, ^{+0.04} _{-0.05}$ \\
&$\sigma$ (eV)                  &$<$185  &    $<$140  & $<$126 \\
&EW              (eV)         &$-22 \, ^{+16} _{-15}$  &  $-21 \pm 11$ & $-31 \, ^{+14} _{-12}$ \\

\hline
\label{tab:orbital}
\end{tabular}
\end{center}
\end{table*}

\subsection{X-ray spectra}
\label{subsect:spectrum}

In order to minimize the effect of any background variations on the
extracted spectra, intervals where the $>$10~keV PN count rate (one
CCD) was $<$1.4~s$^{-1}$ were selected. These selected intervals are
indicated by a thick horizontal line on top of
Fig.~\ref{fig:lightcurve}. For the 2001 spectrum the dip was excluded.
This resulted in exposures of 9.7~ks and 12.5~ks for the 2001 and 2002
observations, respectively.  The spectra were rebinned to oversample
the full width half-maximum of the energy resolution by a factor 3 and
to have additionally a minimum of 20 counts per bin to allow use of
the $\chi^2$ statistic.  In order to account for systematic effects a
2\% uncertainty was added quadratically to each spectral bin.  The
photo-electric absorption cross sections of Morrison \& McCammon
(\cite{mc:83}) are used throughout.  All spectral uncertainties are
given at 90\% confidence. An initial examination of the 2001 and 2002
spectra showed that they are remarkably similar and in what follows we
discuss the results of the (longer) 2002 February observation and
indicate where the 2001 January results differ significantly.

The overall continuum observed in 2002 February was modeled using
absorbed multicolor disk-blackbody and power-law components.  In the
multicolor disk-blackbody (Mitsuda et al. \cite{m:84}; Makishima et
al. \cite{m:86}) $R_{\rm in}$ is the innermost radius of the disk, $i$
the disk inclination angle and $kT_{\rm in}$ the blackbody effective
temperature at radius $R_{\rm in}$.  A power-law was chosen rather
than the Comptonization model used by Iaria et al.~(\cite{i:01}) or
the cutoff power-law used by Smale et al.~(\cite{s:02}) due to the
more limited energy range of the EPIC instrument.  This models the
overall shape of the continuum moderately well and gives a $\chi ^2$
of 392.1 for 234 degrees of freedom (dof).  In contrast to the RXTE
results of Smale et al.~(\cite{s:02}), the disk-blackbody provides
significantly better fits than a blackbody, and no cutoff to the
power-law is needed at $\sim$6~keV.  Examination of the residuals
(middle panels of Fig.~\ref{fig:spectrum}) reveals broad structure
between 1--2~keV.  This is probably the same structure noted by Iaria
et al.~(\cite{i:01}) using BeppoSAX and modeled as an absorption edge
at 1.27~keV and an optical depth, $\tau$, of 0.15.  In order to model
this structure in the PN spectrum, an absorption edge at $1.20 \pm
0.04$~keV with $\tau = 0.051 \pm 0.013$ was added to the model.  This
results in a $\chi^2$ of 346.9 for 232 dof.  However, significant
structured residuals remain.  These are much better accounted for if
the edge is replaced by a Gaussian emission feature with an energy of
$0.93 \, ^{+0.05} _{-0.09}$~keV, a width ($\sigma$) of $175 \pm 65$~eV
and an equivalent width, EW, of $29 \pm 13$~eV to give a $\chi ^2$ of
312.7 for 231 dof.  The F-statistic value of 19.6 for the addition of
the line feature indicates that the probability of such a decrease
occurring by chance is $3 \, 10^{-11}$.  The narrow features at 
1.8~keV and 2.2~keV visible in Fig.~\ref{fig:spectrum} may be due to
an incorrect modeling of the instrumental Si and Au edges.

\begin{figure}
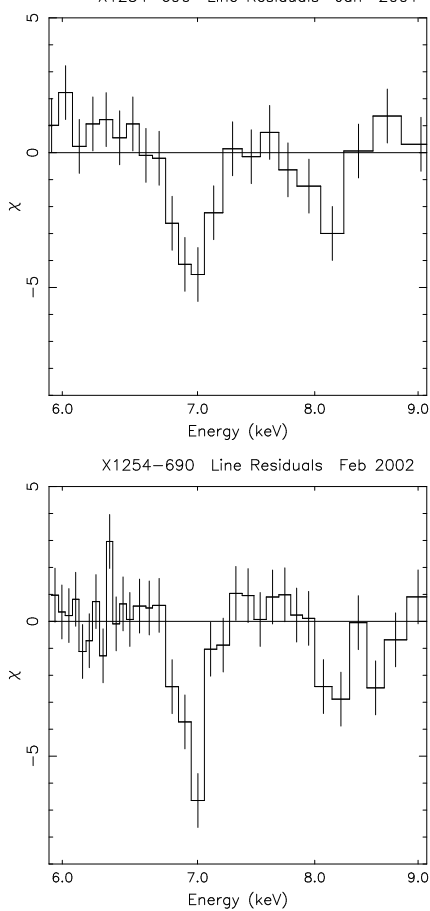

  \hbox{\hspace{1.5cm}
   \includegraphics[width=6.0cm,angle=-90]{H4477F4a.ps}}
   \vspace{0.2cm}
   \hbox{\hspace{1.5cm}
   \includegraphics[width=6.0cm,angle=-90]{H4477F4b.ps}}
  \caption[]{Residuals in the 6.0--9.0~keV energy range
   when the best-fit models given in Table~\ref{tab:spectrum} are
   fit to the PN spectra of the persistent emission  
   and the normalisations of the
   narrow absorption features at 6.96~keV and 8.16~keV are set to zero.}
  \label{fig:dips_res}
\end{figure}

%modif
% addition of reference to fig4 (dips_res)

Examination of the remaining fit residuals in both spectra shows deep
negative residuals at $\sim$7~keV (Fig.~\ref{fig:dips_res}). These were
modeled by a Gaussian absorption line with an energy of $6.96 \pm
0.04$~keV, a width ($\sigma$) of $<$95~eV and an equivalent width (EW)
of $-21 \, ^{+8} _{-5}$~eV.  This results in a $\chi^2$ of 264.8 for
228 dof.  The F-statistic value of 13.7 indicates that the probability
of such a decrease occurring by chance is $3 \, 10^{-8}$.  It is
probable that a second narrow ($\sigma < 80$~eV) absorption feature is
present at an energy of $8.16 \pm 0.06$~keV with an EW of $-16 \,
^{+9} _{-8}$~eV.  Including such a feature in the fit results in a
$\chi^2$ of 244.6 for 225 dof. The F-statistic value of 6.2 indicates
that the probability of such a decrease occurring by chance is only $5
\, 10^{-4}$.  The best-fit energy is close to that of Fe\,{\sc xxvi}
K$\beta$ at 8.26~keV, a feature seen in the ASCA spectrum of
GRS\,1915+105 by Kotani et al.~(\cite{k:00}) and probably in the
XMM-Newton spectrum of X\thinspace1624$-$490 by Parmar et
al.~(\cite{p:02}).  We caution that this is a spectral region where
the EPIC calibration is still relatively uncertain, and so we regard
the presence, and proposed identification, of the feature as probable,
but not confirmed.  The best-fit continuum and line model parameters
for both observation are given in Table~\ref{tab:spectrum}. The 90\%
confidence upper-limit to a narrow Fe\,{\sc xxv} absorption line at
6.70~keV in the 2002 February observation is $-$6~eV.

\subsection{Orbital dependence of the spectral features}
\label{sect:orbital}

%The approximate phases covered are 0.57--0.88, 0.03--0.37 and
%0.37--0.70

In order to investigate whether the properties of the absorption
features depend on orbital phase, the persistent emission interval of
the 2001 observation was divided into three parts of $\sim$3220~s.  To
estimate the phases covered by these intervals, we use a period of
3.88~h and the time 19.25~h of 2001 January 22, corresponding to the
apparent dip center (see Fig.~\ref{fig:lightcurve}, as an arbitrary
reference for phase 0.  The same continuum model as used for the total
persistent spectrum was fit to the 3 PN spectra. In addition, the
Fe\,{\sc xxvi} K$\alpha$ absorption feature is present. The Fe\,{\sc
xxvi} K$\beta$ absorption feature is not significantly detected. The
properties of the Fe\,{\sc xxvi} K$\alpha$ absorption features in the
3 segments are given in Table~\ref{tab:orbital}.  They are all
consistent with those obtained from the total persistent
spectrum. Thus, there does not appear to be any obvious dependence of
the Fe\,{\sc xxvi} K$\alpha$ absorption feature properties on orbital
phase during the persistent emission.

\section{Discussion}
\label{sect:discussion}

We have succesfully modeled the 0.5--10~keV continuum of \src\ during
non-dip intervals using a combination of disk-blackbody and power-law models
together with a broad emission feature at $\sim$0.95~keV and narrow absorption
features at 7.0~keV and 8.2~keV. The nature of the emission feature is uncertain.
The same structure was probably modeled as an absorption edge by Iaria~et al.~(\cite{i:01})
from BeppoSAX data, but the form of this feature clearly depends on the underlying
continuum model chosen. For example, if a blackbody is substituted for the 
disk-blackbody in the 2002 February fits, then the emission feature has an
energy of $0.98 \pm 0.02$~keV and a width of $<$100~eV.Thus while the data
clearly deviate from the chosen underlying continuum in this region, the
exact form of the deviation is uncertain. We note, however, that the properties
of the narrow absorption features are only very weakly dependent on the
overall continuum model chosen.

Narrow X-ray absorption lines were first detected from the
superluminal jet sources GRO\,J1655$-$40 (Ueda et al.~\cite{u:98};
Yamaoka et al.~\cite{y:01}) and GRS\,1915+105 (Kotani et
al.~\cite{k:00}; Lee et al.~\cite{l:02}). ASCA observations of
GRO\,J1655$-$40 revealed the presence of absorption features due to
Fe\,{\sc xxv} and Fe\,{\sc xxvi} which did not show any obvious
dependence of their EWs on orbital phase.  ASCA observations of
GRS\,1915+105 revealed, in addition, absorption features due to
Ca\,{\sc xx}, Ni\,{\sc xxvii} and Ni\,{\sc xxviii}. A recent {\it
Chandra} HETGS observation of this source revealed absorption edges of
Fe, Si, Mg, and S as well as resonant absorption features from
Fe\,{\sc xxv} and Fe\,{\sc xxvi} and possibly Ca\,{\sc xx} (Lee et
al.~\cite{l:02}).  Until recently, it was possible that these
absorption features were peculiar to superluminal jet sources and
related in some way to the jet formation mechanism. With the discovery
of narrow absorption features from the LBXRBs GX\thinspace13+1 (Ueda
et al.~\cite{u:01}), MXB\thinspace1658$-$298 (Sidoli et
al.~\cite{si:01}), X\thinspace1624$-$490 (Parmar et al.~\cite{p:02})
and now \src, this appears not to be the case, and as proposed by
Kotani et al.~(\cite{k:00}) ionized absorption features may be common
characteristics of disk accreting systems.  However, it is interesting
to note that 3 of the above 4 LBXRBs are dipping sources. These are
systems that are viewed from directions close to the plane of their
accretion disks with $i \sim$60--$80\degmark$ (Frank et
al.~\cite{f:87}). Furthermore, GRO\,J1655$-$40 has been observed to
undergo deep absorption dips (Kuulkers et al.~\cite{k:98}) consistent
with observing the source at an inclination angle, $i$, of
$60\degmark$--$75\degmark$ (e.g., Frank et al.~\cite{f:87}). An
inclination of 69.5$\pm$0.3$\degmark$ is independently attributed to
GRO\,J1655$-$40 from optical observations (Orosz \& Bailyn
\cite{1655:orosz97apj}). An inclination of $\sim$70$\degmark$ is
attributed to GRS\,1915+105 assuming that the jets are perpendicular
to the accretion disk (Mirabel \& Rodriguez
\cite{1915:mirabel94nature}). This suggests that inclination angle is
important in determining the strength of these absorption features
which implies that the absorbing material is distributed in a
cylindrical, rather than a spherical geometry, around the compact
object. The azimuthal symmetry is implied by the lack of any orbital
dependence of these features.

The upper-limit to a narrow Fe\,{\sc xxv} absorption feature from
\src\ of $-$6~eV and the Fe\,{\sc xxvi} EW of $-21 \, ^{+8} _{-5}$~eV
implies a ratio of H-like to He-like line EWs of $\approxgt$2.7.
Comparison of this ratio with those from the above LBXRBs shows that
only X\,1624$-$490 has a comparable value of $2.2^{+0.6}_{-2.0}$
(Parmar et al.~\cite{p:02}), whilst for MXB\thinspace1658$-$298 the EW
ratio is 1.3$^{+0.4}_{-0.9}$ (Sidoli et al.~\cite{si:01}) and for
GX\,13+1 it is 1.6$^{+0.3}_{-1.0}$, 1.5$^{+0.4}_{-0.9}$ and
$1.2\pm0.4$ during three different observations (Sidoli et
al.~\cite{si:02}).  This suggests that the material responsible for
the absorption features in \src\ is more strongly ionized than in
MXB\thinspace1658$-$298 and GX\,13+1.  This difference is unlikely to
be caused by observing the central source through hotter material
located closer to the disk since MXB\,1658$-$298 which shows eclipses
and dips and has $i \sim 72$--$90\degmark$ (Cominsky \&
Wood~\cite{cw:89}) has a EW ratio of 1.3$^{+0.4}_{-0.9}$. (However,
the lines from MXB\,1658$-$298 have the highest overall EW). It is
also unlikely to be related to the total source luminosity since
GX\,13+1 with 1--10~keV luminosity of $7\, 10^{37}$~erg~s$^{-1}$ and
an EW ratio of $\sim$1.4 is around a factor 7 more luminous than \src.
One possibility is that this difference is related to the overall size
of the Roche lobe around the compact object into which the accretion
disk must fit. Systems with shorter orbital periods are therefore
expected to have smaller accretion disks. The orbital period of 3.9~hr
for \src\ is signicantly less than for MXB\,1658$-$298 (7.1~hr),
X\,1624$-$490 (21~hr) and GX\,13+1 (593~hr). Since the
photo-ionization parameter, $\xi = L{\rm /n_e \, r^{2}}$, (see e.g.,
Kallman \& McCray~\cite{km:82}) where $L$ is the luminosity of the
ionizing source, $n{\rm _e}$ is the electron density and r is the
distance to the ionizing source, depends on r$^{-2}$, the obscuring
material in smaller systems may be expected to be more photo-ionized
than in larger systems.  This may well explain the high ionization
level of the absorbing material in \src. On the basis of the
properties of the Fe emission feature seen from XB\,1916$-$053, Parmar
et al.~(\cite{p:02}) predicted that this 0.53~hr dip source should
exhibit absorption features. If the high ionization state of the
absorbing material in \src\ is due to its closeness to the central
object, then the same should be true in the case of XB\,1916$-$053.

\begin{acknowledgements}
Based on observations obtained with XMM-Newton, an ESA science mission
with instruments and contributions directly funded by ESA member
states and the USA (NASA).  L. Boirin acknowledges an ESA
Fellowship.

\end{acknowledgements}

\end{document}